# Using smartphones for low-cost robotics

Jose Berengueres-*Member, IEEE*

*Abstract*— **Smartphones and robots can have an adversarial or a symbiotic relationship because they strive to serve overlapping customer needs. While smartphones are prevalent, humanoid robots are not. Even though considerable public and private resources are being invested in developing and commercializing humanoid robots, progress seems stalled and no humanoid robot can be said to be successful with consumers. A part from the obvious engineering differences between humanoids and smartphones, other economic factors influence this situation. On one hand, the product cycle of robots is slower than smartphones. This makes robot computing hardware, (as it with automobile's infotainment systems), perennially outdated when side-by-side to a smartphone. On the other hand, the incentives to develop Apps for smartphones are high and they are comparatively low for robot platforms. Here, we point to how smartphones could be used to lower hardware cost and foster robot app development.**

## I. INTRODUCTION

While smartphones are prevalent, humanoid robots are not (yet). Even though considerable public and private resources are being invested in developing and commercializing humanoid robots, no robot project seems to equal the success of smartphones. Here, we address some possible causes from an engineering point of view but also from an economic point of view and we put them in relation to smartphones. This survey is sparked by the lessons learnt in developing a humanoid robot that uses two off-the-shelf smartphones as compute units instead of dedicated hardware [1]. Following, we first identify a persistent hardware performance gap between smartphones and robots and we attribute it to three root causes: (i) software network effects [2], (ii) hardware manufacturing economies of scale [3] and, (iii) the *product leadership* [4] economic advantage that smartphones hold over robots. We then address the main roadblocks that affect the success of consumer facing robot projects, and finally, we layout critical cost-reduction benefits of using a mobile OS as a development platform.

### A. Affordable robots

Despite the successes of Amazon's *Alexa*, *Google Home* and iRobot's *Roomba*, the future of consumer home robotics might have humanoid shape. There are three main reasons that point in this direction. First, is that humans prefer to talk to things that look human, if we are not too deep into the uncanny valley [5,6]. A second, (somehow controversial) argument is that, as Isaac Asimov pointed in the 70's, we have so many tools that are designed for and by humans that a robot any other shape will not be able to fully utilize them. A third factor is related to safety. Two standards regulate the safety of collaborative and industrial robots: the ISO 15066 and the ISO 10218-1:2011 [7-8]. However, because safety improves

This work was supported in part by the UPAR G25002 grant.

as communication improves, and anthropomorphic is an optimal choice for communication [9], then in a wide range of user facing applications, a humanoid UI seems an optimal choice. As such, a few companies like Rethink robotics (maker of the *Sawyer* robot [10]), Softbank (maker of *Nao* and *Pepper* robots) and Hanson Robotics (maker of *Sophia* robot) have leveraged the humanoid shape to carve a niche in their respective markets. However, none of them has reached the threshold of mass-production and, to date, all humanoid robots sold are still assembled by hand or in small batches. Meanwhile, their current cost-performance is not attractive enough for any such robot to be adopted by the middle-class family. In other words, the humanoid robots available-to-date are not affordable.

### B. Robot revolution, slower than the automobile

This situation is reminiscent of the dawn of the automobile industry. Before the Ford model-T appeared in 1908, a typical automobile was priced at 2-years of annual income of the average American family. It was only when the Ford-T came to market with an "affordable" price tag of 6-months annual income, that mass adoption took off [11]. Interestingly, since the patent of the gasoline engine by Karl Benz in 1879 to the 1908's Ford-T, **39** years elapsed. In robotics, since the first industrial robot company (Unimation) was founded in 1956, Connecticut, to 2018, **62** years have elapsed. This might lead some historians to question when will mass market robotics happen. To clarify this, let's identify non-engineering roadblocks standing between projects such as *Romo, Jibo, Nao, Asimo, Sophia* and so on, and mass-market success.

### C. Towards mass adoption

Apart from the formidable scientific challenges that roboticists and engineers are tackling to build humanoid robots (biped walking, navigation, manipulation, cognition, conversation, computer vision, real time software, safety). And, despite, the formidable and accelerating progress achieved in the recent years showcased in events such as the DARPA challenge, Amazon pick up challenge, Robocup, and annual conferences such as IEEE IROS and IEEE ICRA. Here, we want to point out to other non-engineering reasons that might be worth considering regarding affordability and popularization. One straightforward factor, among many, of why humanoid robots are not a consumer success yet is *low comparative utility*. For example, Softbank's impressive *Pepper* robot is currently similar to a smartphone in terms of end-user utility. This is due to the fact that there are not as many App developers developing Apps for Pepper as there are for smartphones. This, is a consequence of three economic forces at play. The network effects in software development in smartphones, the law of economies of scale in hardware manufacturing, and the product leadership [6] of smartphones over humanoid robots. Let's analyze how these three forces affect humanoid robotics from a product point of view.

J. B., is with CS Dept. at CIT, UAE University, Al Ain, AD, 17551 UAE. (tel: +971553519573, e-mail: jose@uaeu.ac.ae).

## II. Comparison robot vs. smartphone

Of the robots mentioned earlier, the closest example to a commercial success is perhaps Softbank's *Pepper* and her sister *Nao*, both retailing for about $20,000 and $10,000 respectively. It is estimated that Softbank has sold approximately 1k and 10k units respectively, and while they are an engineering marvel of integration of technology, their abilities are mainly limited by the processing power of an Intel Atom CPU (2MB/512k L2 cache memory) and an onboard battery that is not designed to support new power hungry hardware. (e.g. Nvidia Jetson, Nvidia-Tegra, and so on).

### A. The perennial robot-2-smartphone gap

These drawbacks are compounded by the fact that smartphones' performance seems perennially ahead of that of any other robot. They are also more affordable. To illustrate this, Fig.1 shows how six generations of a best-selling smartphone compare to three generations of a best-selling humanoid robot. Fig.1 Y-axis is cache memory of the CPU. It is taken as an indicator of overall processing performance, as cache size is correlated to the maximum computation complexity that a system can undertake in parallel. For example, a humanoid robot must typically run in parallel and continuously a wake word detection loop, a conversation loop and a computer vision loop [1]. From [1] we know that this is not possible to run smoothly with only 5MB of cache memory on a single commercial off-the-shelf smartphone equipped with a general-purpose CPU.

### B. Product cycle advantage

From Fig.1 we note an additional important point. Smartphones have a quicker product cycle than humanoid robots. Smartphones have a product-cycle close to 18 months. On the other hand, the Nao robot for instance, has a cycle close to that of automobiles with a typical 3-7 years between hardware updates. (This gap is not only exclusive between robots and smartphones. The same situation arises when comparing with infotainment systems embedded in cars). In the robot proposed in [1], two smartphones are used in tandem as compute units. The combined L cache memory is 10MB. This is 20 times more than some Pepper and Nao models (512kB model). If we factor in the cost $8,000.00[1] vs. $20,000.00, on a kB per dollar spent, the gap is x50. Fig.2 compares the same datapoints of Fig.1, but on a kB of L cache memory per dollar spent. This ***robot-2-smartphone*** gap is present in many other key hardware elements: camera image quality in low light, screen quality and so on. Finally, this gap between robot and smartphones seems consistent over time. As a note, even after Softbank (at the time Aldebaran) switched from the Geode CPU to Intel Atom chips in 2014, the gap remained unchanged. Table 1 compares more specs between the 2-smartphone powered humanoid of [1] and the top seller Pepper.

### C. Developers incentives

The second roadblock between robot projects and mass market is the lack of Apps that exploit humanoid robot's capabilities. Why is this? Various estimates indicate that there are about 19 million software developers in the world, and that of those about 50% are dedicated to making apps for mobile devices [12,13]. In addition, there are about 400,000 App publishers combining Play Store, App Store and Amazon data. The global mobile App gross revenue in 2016 was $88 billion and is expected to reach 189bn by 2020.

TABLE I.    PLATFORM COMPARISON HW

|  | 2 x Smartphone powered robot [1] | Softbank Pepper |
|---|---|---|
| Year model | 2018 | 2017 |
| OS | iOS | NAOqi /ROS |
| Main HW | iPhone 6+ iPad (2013) Raspberi Pi | Pepper robot |
| CPU | A8 + A7 | Atom Z350 |
| Available L cache (L1,L2,L3,L4) | 10 MB | 2MB/512 kB |
| Max camera pixels | 8 MP | 5[a] MP |
| Cost | $8,000.00 | $20,000.00 |
| DoF | 13 | 20 |
| cost /DoF | ~$500 | ~$1000 |

a. OV5640 Pepper camera model 2017.

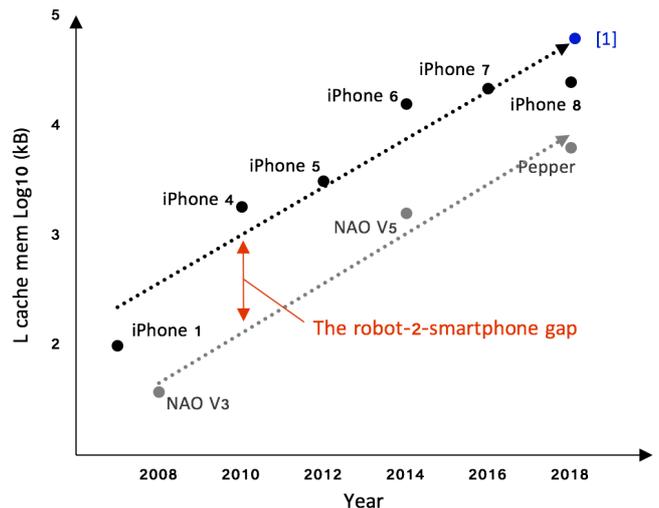

Fig. 1 Six generations of a best-selling smartphone are compared to the best-selling humanoid robot. Using the CPUs total L cache memory size as an indicator of performance, we can see that there is a "perennial" 10x gap between robot hardware and smartphone hardware despite the former being costlier. Data source: manufacturer, ARM, Intel.

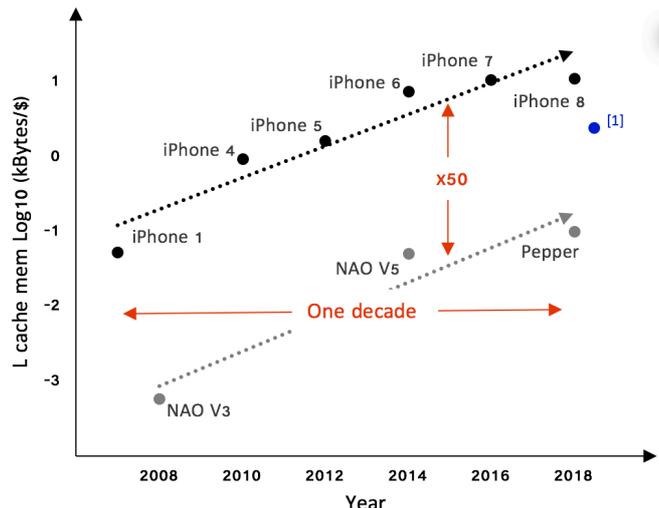

Fig. 2 On a per dollar spent basis the robot-2-spartphone gap spans one decade. This makes robots unattractive to smartphones when compared side by side.

Of those revenues, about **13%** is from ads, **50%** from in-app purchases (aka micro transactions), and the rest from apps purchases. The largest mobile App store in the world, has 2.4 million apps available [12]. These Apps emerged organically from an incentive structure put forward by the App store owners that allowed developers: (i) to develop efficiently thanks to a powerful SDK; (ii) to monetize the investment in Apps via a curated store that is trusted by users. With a 30-70 revenue-split revenue sharing structure between the store owner (in the case of iOS, Apple) and the developers, these monetization schemes have been very successful with developers. This, coupled with the *product leadership* [4] of smartphone hardware, created the virtuous cycle that we know as the App boom or the **App Economy** [13, 14]. However, when we look at Apps available specifically for robots, the most successful robot-App store today has about 1,000 Apps (a gap of 2000x with Android) and only about 10k potential user-owners (corresponding to the total units sold of Pepper and Nao robots). However, the main incentive to attract developers to the platform a SDK native monetization mechanism and an App store with a critical mass of users is not available [15].

### D. Number of Apps available

Fig.3 compares quantitatively various robot platforms and the number of Apps available for each one. We take the number of Apps (github projects when not available) as a proxy for utility. Note that in Fig.3 the two Android and iOS based robots (in blue) are matched to the Apps available to their corresponding OS platform, this is not entirely accurate. However, from the experience of [1], we recorded that it costs approximately an additional 30% of code to convert an App initially developed for mobile to a humanoid robot platform such as [1] or the Sony Xperia G1209. This assumes that the same developer team performs the translation (*humanoid-ization*) from mobile App version to humanoid version. This process consists mainly about adding body language, arm gesture and voice recognition dialogue loops to the existing functionality. In other words, IU/UX enhancements. Fig.4 is a reprint of the lines of code tally from the *humanoid-ization* case from [1].

### E. SDK productivity

Finally, another important factor that has contributed equally to the economic and creative success of the App Economy was not only the App store itself, but the investment by App store owners in quality SDKs. An efficient SDK is the one that lets developers focus on building their App rather than on necessary, but non-core surrounding elements to their App (typically UI functions, payment functions and so on). Non-coincidentally, the improvement of the SDKs brought also the standardization of design elements (UI components). This standardization, in Apple consisting of a set of guidelines called Human Interface Guidelines [16] and accompanying component libraries and in Android known as the Material Design among other bits and pieces, freed developers of building common UI elements (such as buttons, sliders, menus, animations) from scratch. This, standardized the look and feel of the Apps and improved the user experience across the board while dramatically reducing developer billed-time.

### F. SDK

In the robotics world, there have also been extremely valuable standardization drives such as ROS (Robot Operating System). ROS is open source and is currently the most used robot OS. While ROS helped standardize how robot software interacts with the myriad of underlying robot hardware systems available, saving precious software development time to roboticists by helping them reuse code and libraries, it was never intended to address the same problems that mobile SDK addresses. These are basically making monetization, and UI coding easy to the developer. In other words, the ROS building blocks are excellent in abstracting robot hardware and at addressing robot-specific functions such as navigation, servo control or inverse kinematics, but ROS was never intended for commercial app development. Such functionality might not be core to the robot but is indispensable to develop a consumer ready robot-Apps.

Finally, an additional point to consider is the know-how and libraries available for each SDK. The more know-how available the easier to develop useful Apps. We take the number of published projects and apps as a proxy for accumulated know-how. A search on github reveals **36,338** ROS software projects while for Android there are **698,487** projects uploaded, a 20x gap. Table 2 shows additional details.

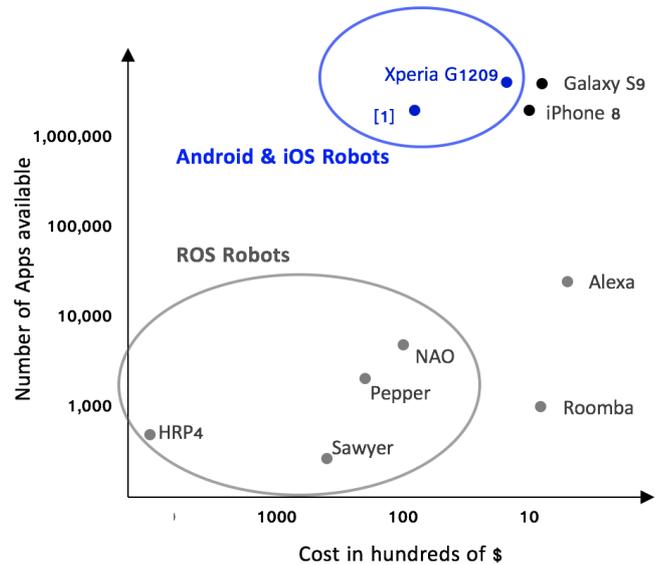

*Fig. 3 Robots running on smartphone platforms can run apps initially designed for mobiles and hence can benefit natively from leading SDKs and know-how. Source: github, Google play, Apple store, Amazon.*

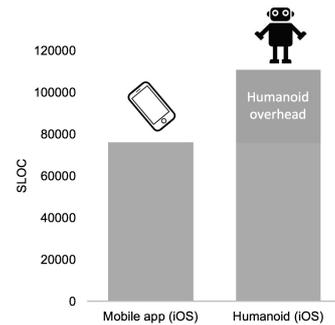

*Fig. 4 Building a humanoid version of the mobile App only took an additional 31% SLOCs. 90% of the mobile App code was reused. Reprinted from [1]*

TABLE II.    PLATFORM COMPARISON SW

| metric | Mobile | Softbank Pepper store |
|---|---|---|
| OS | iOS | NAOqi /ROS |
| Apps available | 2.4M | ~1000 |
| Monetization mechanism | Y | N |
| Developer Revenues | $86bn[a] | $0.00 |
| Hits in github | **232k** | **4k** |

a.    Includes all mobile app revenues 2016.

## III. CONCLUSION

Smartphones and robots can have an adversarial or a symbiotic relationship because they strive to serve overlapping customer needs. The leading SDKs today are Android and iOS. The largest pool of developers today is skilled in mobile SDKs and not robot specific SDKs. History shows that the application store is as important as the quality of the SDK to have a functioning App ecosystem. Without a carefully thought incentive structure developers cannot invest their time and resources in developing robot-Apps. And without Apps the potential utility of the hardware and other robot specific skills such as navigation, manipulation cannot be fully enjoyed by the end user. In addition, current robot-SDKs where not designed to produce consumer apps such as mobile SDK were in the first place. All this poses challenges to the consecution of mass-market humanoid robots. The SDK combined with the performance gap pointed at suggests that if robots are to become as popular as smartphones, a low-cost route might very well be to leverage the smartphone hardware, resources and existing know-how. Since popular Apps already exist in mobile and tablet versions, customizing an existing App for an extra device (in this case a humanoid robot) seems a logical path towards producing robot-Apps efficiently, a key step towards low-cost humanoid robots.

### ACKNOWLEDGMENT

We would like to thank Kuka Robotics AG staff at Hannover Messe 2018, Kenjiro Tadakuma, Tohoku Daigaku, Lojain Jibawi and Sawsan Said, and Franka Emika staff for fruitful discussions.

### REFERENCES

[1] Jibawi, L., Said, S., Tadakuma, K., & Berengueres, J. (2018). Smartphone-based Home Robotics. arXiv preprint arXiv:1803.02122.

[2] Katz, M. L., & Shapiro, C. (1994). Systems competition and network effects. Journal of economic perspectives, 8(2), 93-115.

[3] Mendelson, H. (1987). Economies of scale in computing: Grosch's law revisited. Communications of the ACM, 30(12), 1066-1072.

[4] Cooper, R. G. (1999). Product leadership: creating and launching superior new products. Basic Books.

[5] Epley N, Waytz A, Cacioppo JT. On seeing human: a three-factor theory of anthropomorphism. Psychological review. 2007 Oct;114(4):864.

[6] Mori, M., MacDorman, K. F., & Kageki, N. (2012). The uncanny valley [from the field]. IEEE Robotics & Automation Magazine, 19(2), 98-100.

[7] Rosenstrauch, M. J., & Krüger, J. (2017, April). Safe human-robot-collaboration-introduction and experiment using ISO/TS 15066. In Control, Automation and Robotics (ICCAR), 2017 3rd International Conference on (pp. 740-744). IEEE.

[8] Fryman, J., & Matthias, B. (2012, May). Safety of industrial robots: From conventional to collaborative applications. In Robotics; Proceedings of ROBOTIK 2012; 7th German Conference on (pp. 1-5). VDE.

[9] Duffy, B. R. (2003). Anthropomorphism and the social robot. Robotics and autonomous systems, 42(3-4), 177-190.

[10] Guizzo E, Ackerman E. How rethink robotics built its new baxter robot worker. IEEE spectrum. 2012 Sep 18:18.

[11] Kim, W. C., & Mauborgne, R. (2004). Blue ocean strategy. If you read nothing else on strategy, read thesebest-selling articles., 71.

[12] Martin W, Sarro F, Jia Y, Zhang Y, Harman M. A survey of app store analysis for software engineering. IEEE transactions on software engineering. 2017 Sep 1;43(9):817-47.

[13] Businesofapps.com, ' There are 12 million mobile developers worldwide, and nearly half develop for Android first ', 2017. [Online]. Available: http://www.businessofapps.com/12-million-mobile-developers-worldwide-nearly-half-develop-android-first/ . [Accessed: 23- Jan- 2018].

[14] Basole, R. C., & Karla, J. (2011). On the evolution of mobile platform ecosystem structure and strategy. Business & Information Systems Engineering, 3(5), 313.

[15] Aldebaranrobotics.com, 'SoftBank Robotics Store Apps', 2017. [Online]. Available: https://cloud.aldebaran-robotics.com/ . [Accessed: 23- Jan- 2018].

[16] Apple.com, ' Human Interface Guidelines ', 2017. [Online]. Available: https://developer.apple.com/design/ . [Accessed: 23- Jan-2018].

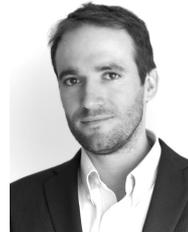

**J. Berengueres** was born in Zaragoza, Spain, in 1976. He received a M.S. degree in international development engineering from the Tokyo Institute of Technology, Japan, in 2005 and the Ph.D. degree in engineering on the topic of bioinspired robots, in 2007.

From 2007 to 2008, he worked at the CMS experiment at CERN. From 2009 to 2010 he participated in two software startups. Since 2011, he has been faculty at the CS Department, College of IT, UAE University, Al Ain, Abu Dhabi. He is chair of the UAE EMBS and RAS chapter. His research interests include design thinking methods, applied data science, and robotics.

Dr. Jose research has received various awards such as the MRS Society 2nd best video award, in 2007, the HRI 2nd best video award in 2013, and the M.B.R Al Maktoum distinguished research award in 2017.